\documentclass[fleqn,twoside]{article}
\usepackage{espcrc2}

\usepackage{graphicx}
\usepackage[figuresright]{rotating}

\newcommand{\AmS}{{\protect\the\textfont2
  A\kern-.1667em\lower.5ex\hbox{M}\kern-.125emS}}

\title{Active Galactic Nuclei, Radio Jets and Acceleration of UHECRs}

\author{S. Massaglia\address[]{Dipartimento di Fisica Generale dell'Universit\`a, \\ 
        Via Pietro Giuria 1, 10125 Torino, Italy	}%
        }
       
\begin{document}

\begin{abstract}
We present the general properties of the Active Galactic Nuclei (AGNs) and discuss the origin
and structure of jets that are associated to a fraction of these objects. We then
we address the problems of particle acceleration at highly relativistic energies and 
set limits on the luminosity of AGN jets for being origin of UHECRs.
\vspace{1pc}
\end{abstract}

\maketitle

\section{INTRODUCTION}

\begin{figure}[htb]
\vspace{9pt}
\includegraphics[width=0.45\textwidth]{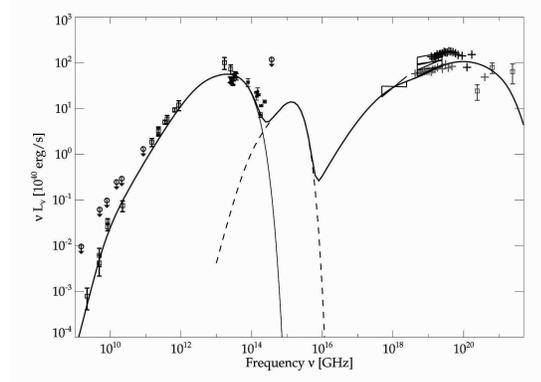}
\caption{Spectral Energy Distribution of the Centaurus A core.
The continuous line is a synchrotron plus Synchrotron Self-Compton (SSC)
model \cite{Prie07}.}
\label{fig:1}
\end{figure}

Most of the galaxies of the Local Universe shine by effect of stellar and interstellar gas
emissions, predominantly in the optical band. Typically, the emitted spectra are characterized by stellar
absorption lines and emission lines by HII regions. These sources are called {\it Normal Galaxies}.
A small fraction of galaxies, about $1 \%$, do not follow this behavior. In fact, they show strong and broad 
emission lines, consistent with velocity dispersion of the emitting gas attaining several thousand of 
kilometers per second. Most remarkably, the non-thermal emission coming from a 
central nucleus, with size $\sim 10^{-2}$ pc, dominates over the thermal one coming from stars and interstellar gas and extends
well beyond the optical band, from radio to gamma rays (Fig.~\ref{fig:1}). 
These sources are called {\it Active Galaxies} and host {\it Active Galactic Nuclei} (AGNs) at their centers.
\begin{table*}[htb]
\caption{}
\label{tab:1}
\newcommand{\m}{\hphantom{$-$}}
\newcommand{\cc}[1]{\multicolumn{1}{c}{#1}}
\renewcommand{\tabcolsep}{2pc} 
\renewcommand{\arraystretch}{1.2} 
\begin{tabular}{@{}ll}
\hline
Radio quiet AGNs & Radio loud AGNs  \\
\hline
Seyfert I galaxies (Sey 1) (BLR, $\sigma \sim 10^4$ km \ s$^{-1}$) & Radio galaxies\\
Seyfert II galaxies (Sey 2) (NLR, $\sigma \leq 10^3$ km \ s$^{-1}$) & Radio quasars\\
Radio quiet quasars (QSOs) & BL Lac Objects\\
~  &  Optically Violent Variables (OVVs)\\
\hline
\end{tabular}\\[2pt]
\end{table*}
AGNs are not all the same, on the contrary they can be extremely different in their properties.
It resulted convenient to classify AGNs according to their radio power, in fact they can be separated into
two distinct classes: Radio Quiet and Radio Loud AGNs. Typically, the luminosity in the
GHz band of radio loud AGNs exceeds the radio quiet ones by about three orders of magnitudes.
Observationally, Seyfert 1
show broad emission lines in the spectrum (Broad Line Region, BLR, with velocity
dispersion $\sigma \sim 10^4$ km \ s$^{-1}$), while for Seyfert 2 the
spectral lines are narrower (Narrow Line Region, NLR, $\sigma \leq 10^3$ km \ s$^{-1}$).
Table~\ref{tab:1} summarizes the objects belonging to the two classes. 

\begin{figure}[htb]
\includegraphics[width=0.45\textwidth]{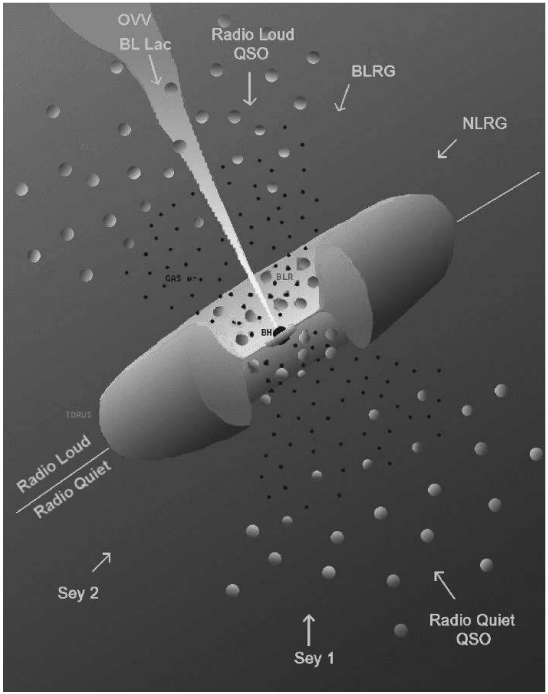}
\caption{The unified AGN model}
\label{fig:2}       
\end{figure}

This complex ``zoology" has been interpreted by \cite{Urry95} within the so called Unified Model
for AGNs. According to this picture (Fig. \ref{fig:2}), radio quiet AGNs have no jets (or very weak ones)
while the radio loud ones have jets; the angle between the line of sight and the plane of the
obscuring torus determines the observed properties of AGNs. We will discuss the main properties of radio
loud AGNs, in particular of the radio galaxies, since the presence of a jet may be crucial for accelerating
cosmic ray at the highest energies.

In Section 2 we review the main properties of radio galaxies and in Section 3 we will discuss the
particle acceleration from AGN jets. The main outcomes are summarized in Section 4.

\section{THE RADIO GALAXIES}

Radio galaxies are seen in the radio band emitting power-law spectra,
indicative of synchrotron emission with typical spectral index $\alpha \sim 0.5$. Moreover, 
jets and hot-spots of some bright sources are seen in the optical and X-ray bands as well.
The emission mechanism at these higher frequencies is again continuum and it may be
synchrotron or Synchrotron-Self-Compton. Direct observations
of radio galaxies give us: i) the radio luminosity  $\sim 10^{39}-10^{44} \ {\rm ergs \ s}^{-1}$;
ii) the size, from a few kiloparsec to some megaparsec; iii) the morphological brightness distribution,
and iv) the polarization degree of the radio emission. We can then derive, by indirect means, the main physical
parameters such as the life timescale, $10^7-10^8$ ys, the mean magnetic field, $10-10^3 \ \mu$G, and the
kinetic power, $10^{42}-10^{47} \ {\rm ergs \ s}^{-1}$. The values of the jet main physical parameters, 
such as jet velocity, density and composition, are still under debate after many decades of investigations. 
The reason for these uncertainties in constraining the basic physical parameters
is due to the very nature of the radiation emission which is
typically non-thermal continuum, i.e. the absence of any lines in the radiation spectrum
\cite{Mass03}. Therefore, the value of the magnetic field is derived by the minimum energy assumption,
the jet kinetic power by the work done against the ambient to evacuate cavities for accommodating the radio lobes,
the jet velocity by the radiative flux contrast of the approaching to receding jets and by proper motion 
observations, and the jet density by comparison between observed morphologies and the outcome of numerical
simulations.

\subsection{The Fanaroff-Riley Classification}

\begin{figure}[htb]
\includegraphics[width=0.5\textwidth]{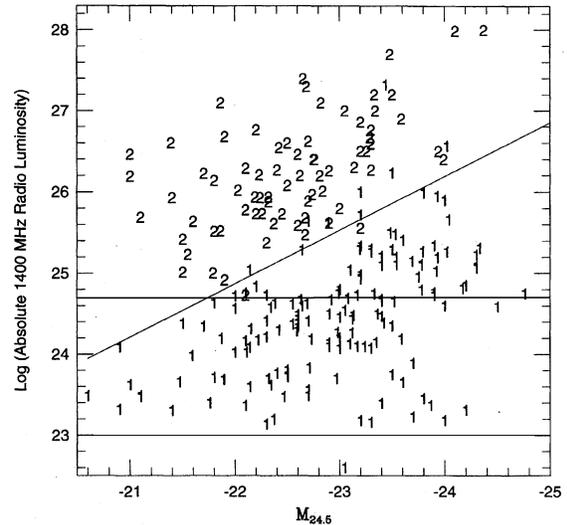}
\caption{Diagram of a sample of FRI/II objects (from \cite{Ledl96}), where
the bottom horizontal line (thin solid line) is the limiting power for radio-loud AGN, while the top one
(thick solid line) is the limiting power for acceleration of UHECRs with the efficiency $\eta=1$ (cf. Eq. \ref{eq:lnu1}).}
\label{fig:ledlow}       
\end{figure}

Historically, the extragalactic radio sources have been classified into two
categories \cite{FR74} based upon their radio morphology: a
first class of objects, preferentially found in rich clusters and hosted by
weak-lined galaxies, shows jet-dominated emission and two-sided jets and  was named FR I; 
a second one, found isolated or in poor groups and
hosted by strong emission-line galaxies, presents lobe-dominated emission and one-sided jets and was 
called FR II (or ``classical doubles").
Besides morphology, FR I and FR II radio sources were discriminated in power
as well: objects below $\sim 2 \times 10^{25} h_ {100}^2$ W Hz$^{-1}$ str$^{-1}$  were typically
referred as FR I sources. A perhaps
more illuminating criterion has been found by \cite{Ledl96} who 
plotted the radio luminosity at $1.4$ GHz
against the optical absolute  magnitude of the host galaxy:
they found the bordering line of FR I to
FR II regions correlating as $L_R \propto L_ {opt}^{1.7} $ (Fig. \ref{fig:ledlow}), i.e. in a luminous 
galaxy more radio power is required to form a FR II radio sources.  This correlation is important 
since it can be interpreted as an
indication that the environment may play  a crucial role in determining the 
source structure. The above argument yields the basic question of the origin of FR I/FR II 
dichotomy, whether intrinsic or ambient driven \cite{Mass03}. We note as well that there are no sources below 
a luminosity of $10^{23}$ W Hz$^{-1}$ (at $1.4$ GHz).

\section{COSMIC RAY ACCELERATION}
The problem of the sites of cosmic ray acceleration, after many decades of experimental
and theoretical investigations, is not yet solved. Concerning the 
cosmic rays in the highest-energy region, above 
 $E=3 \times 10^{18}$ eV, there is a common, even though not unanimous, agreement
about an extragalactic origin. Several mechanisms and classes of astrophysical objects
have been proposed in the literature
\cite{Bier97,NW2000}. Among the proposed candidates for the origin of the
highest energy particles the are: i) neutron stars and other compact object
\cite{Hillas84}; ii) the sources of gamma-ray bursts \cite{Milg95,Vietri95}; iii) large-scale shocks
due to merging galaxies \cite{Cesrky93} or in accretion shocks in clusters of galaxies \cite{Kang96};
iv) the active galactic nuclei \cite{Prot92}; v) the hot-spots of FR II radio galaxies 
\cite{Bier87,Rach93}; vi) the AGN jets by inductive particle acceleration \cite{Bland00,Lyut07}.  

Considering the hypothesis that UHECRs originates from AGNs, and in particular from
the hot-spots of FR II sources via Fermi I acceleration, 
\cite{Mass07} has shown that 15 FR IIs only are located
within 130 Mpc from us, distance set by the GZK cutoff. Moreover, \cite{Lem08} argues that
the FR II sources (NGC4261, PKS1343-60), closest to the highest energy event detected by the Pierre Auger
Experiment, were $\sim 30^\circ$ away from its position. However, more statistics is required
before ruling out this possibility.

On the other hand, UHECRs can come from AGN jets themselves, as pointed out by \cite{Bland00,Lyut07},
 by increasing the energy of protons from $E \le 10^{18}$ eV up to
 $E > 10^{19}$ eV by electric fields induced in sheared, relativistic magnetized jets, whenever:

\begin{eqnarray}
\vec \varepsilon=\nabla \Phi=-\frac{1}{c} \vec v \times \vec B \rightarrow \Delta \Phi =\frac{1}{c} 
\vec B \cdot \left( \nabla \times \vec v
\right) < 0 
\nonumber
\end{eqnarray}
where $\vec \varepsilon, \ \vec B, \ \vec v$ are the electric and magnetic fields and jet velocity, respectively, and
$\Phi$ the electric potential. The resulting particle spectrum becomes $\propto E^{-2}$ asymptotically.

Sheared jets, needed for this kind of acceleration, are actually invoked for interpreting the limb-brightening of
radio jets at parsec scales. One must recall that, while FR I jets are non-relativistic at kiloparsec scales
and FR II ones are relativistic, VLBI observations of radio galaxies \cite{Giov01} have shown that, 
albeit these two kind of sources have
different radio power and kiloparsec scale morphology, they appear similar at parsec scales where the
jet bulk Lorentz factor is typically $\Gamma = 3-10$ for both classes. Moreover, as
said before, about ten FR I sources, e.g. M87 \cite{Kov07}
and B2 1144+35 \cite{Giov07}, observable in the radio band at the VLBI show limb-brightened emission. 
This limb brightening can be interpreted assuming the presence in the jet
of a central spine with high Lorentz factor, $\Gamma =5-10$, and an outer layer with $\Gamma \simeq 2$.
The synchrotron emission from the central spine would be deboosted when viewed at angles larger than
about $30^\circ$.
The problem of the interaction of relativistic, fluid jets with the
ambient was studied numerically, in three spatial dimensions, by \cite{Ros08}; they
followed the onset and nonlinear growth of unstable modes in the jets while propagating into a uniform medium. 
The unstable perturbations caused mixing and
deceleration of the jet, processes controlled by the initial jet Mach number $M$ and ambient-to-jet density ratio
$\eta$, with the jet Lorentz factor kept fixed to 10. 
They found that very light jets were substantially decelerated, and the jet deceleration was not uniform, 
but more effective at the outer layers leaving a central ``spine" traveling at higher Lorentz factor. 
Thus, FR I jets as well may meet the conditions to be source of UHECRs via inductive particle acceleration.
One can then ask the same question of \cite{Mass07}: how many FR I sources, potential accelerators of 
cosmic rays, can be found within $100$ Mpc from Earth? The radio luminosity function (RLF) of a sample 
of radio-loud
AGNs from the $1.4$ GHz NRAO VLA Sky Survey was obtained by \cite{Mau07}. If one extracts from
this RLF the number of objects, with luminosity exceeding $10^{23}$ W Hz$^{-1}$, expected within
$100$ Mpc from us one finds about $100$ sources, that should be identified and their positions 
possibly correlated to the arrival directions of Pierre Auger events.

\subsection{A constraint to the jet luminosity}

Independently of the particular acceleration mechanism considered,
not all radio-loud AGNs can effectively accelerate UHECRs. Cosmic rays must be contained
into the acceleration region in order to attain the maximum energies \cite{Hillas84},
in other words the acceleration timescale must not exceed the dynamical time
available for the acceleration. This condition reads, in the frame comoving with the jet:
\begin{equation}
t_{\mathrm {acc}} \le t_{\mathrm {dyn}}
\label{eq:cond}
\end{equation}
with 
\begin{equation}
t_{\mathrm {acc}} = \eta^{-1} t_{\mathrm L} \;, \;\; t_{\mathrm {dyn}}= \frac{R}{\beta \Gamma c}
\end{equation}
where $\eta\leq 1$ is an efficiency parameter that is appropriate for the actual acceleration mechanism, 
$R$ the jet longitudinal coordinate,
$t_{\mathrm {L}}=E/ZeBc$ is the Larmor time in the jet magnetic field, $\Gamma$ the jet Lorentz
factor and $\beta \ c$ its velocity. Applying condition
(\ref{eq:cond}), one derives the maximum energy achievable, in the observer's frame \cite{Lem08}:
\begin{equation}
E_{\mathrm {obs}} \le \eta Z e BR \beta^{-1}
\label{eq:eobs}
\end{equation}
This quantity must be related with the jet magnetic field intensity. To the jet total power contribute
both kinetic and magnetic (Poynting) luminosities:
\begin{equation}
L_{\mathrm {tot}} = L_{\mathrm {kin}} + L_{\mathrm {B}}
\label{eq:tot}
\end{equation}
with
\begin{equation}
 L_{\mathrm {kin}} = 2 \pi R^2 \theta^2 (n_0 m c^2) \Gamma(\Gamma-1) \beta c 
\label{eq:kin}
\end{equation}
and
\begin{equation}
 L_{\mathrm {B}} = 2 \pi R^2 \theta^2 \frac{B^2}{8\pi} \Gamma^2 \beta c 
\label{eq:mag}
\end{equation}
where $\theta$ is the jet opening angle and $n_0$ the jet density in the comoving frame.
Since magnetic fields are likely to play a crucial role in the
acceleration process, one may think that the main constituent of jets, close to its origin, is not 
mass but fields in form of Poynting-dominated beams. However, \cite{Sik05} argued
that, even though jets could be Poynting-dominated at the origin, observational
data imply that they become kinetically dominated beyond about 1,000 gravitational radii
from the central acceleration region, with at its center a supermassive black-hole of
$\sim 10^{8-10}$ M$_\odot$. Furthermore, \cite{Gian06} discussed the role of kink instability in
Poynting-flux dominated jets and find that the Poynting flux dissipates and the jet
becomes kinetically-dominated, again at about 1,000 gravitational radii. Thus it is therefore
reasonable to assume that if close to the AGN $L_{\mathrm {tot}} \approx L_{\mathrm {B}}$,
at the kiloparsec scale $L_{\mathrm {tot}} \approx L_{\mathrm {kin}} $.

\begin{figure}[htb]
\includegraphics[width=0.45\textwidth]{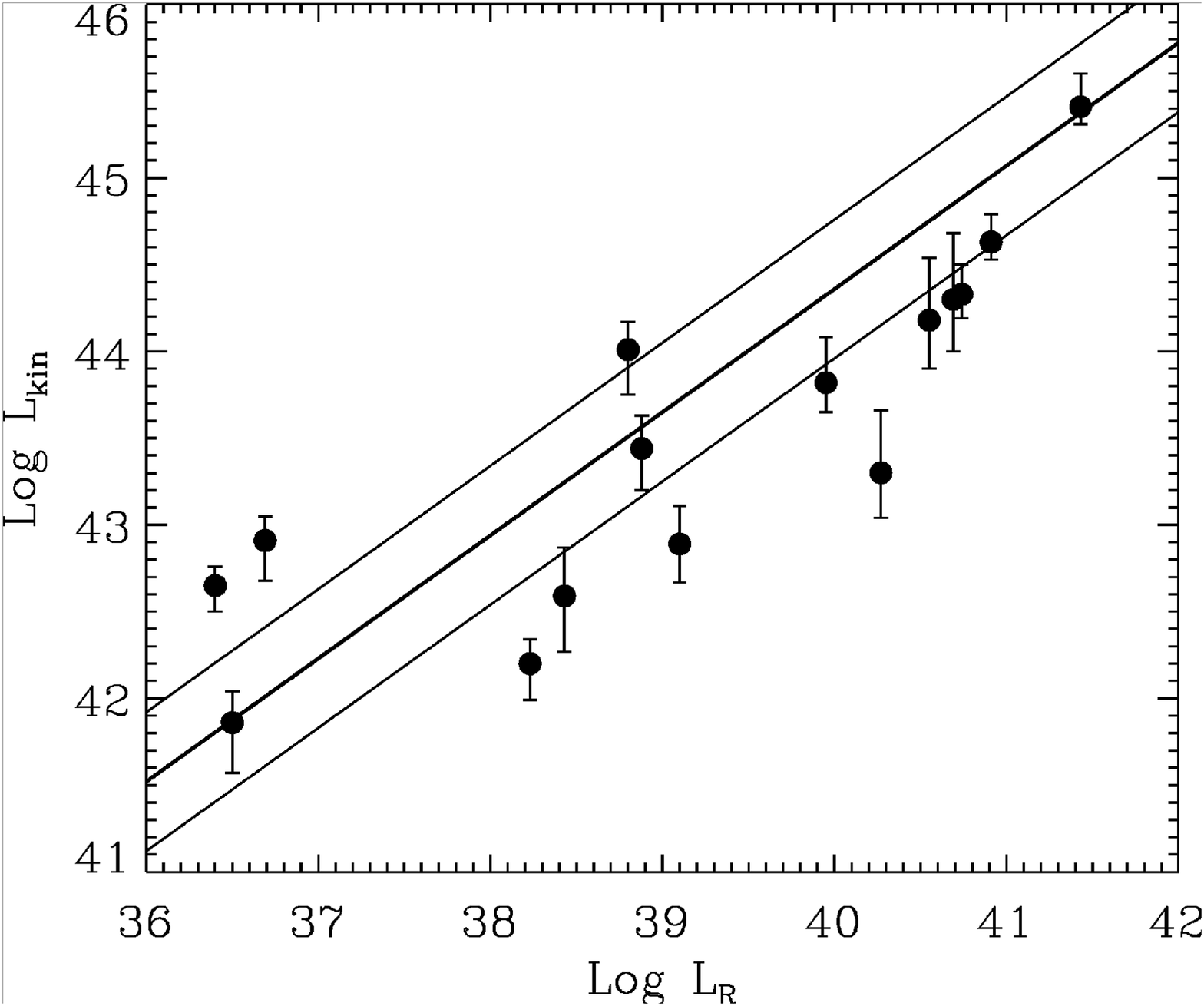}
\caption{Plot of the kinetic power against the core radio power at 5 GHz (from \cite{Hein07}).}
\label{fig:heinz}       
\end{figure}

Combining Eq. \ref{eq:eobs} with Eq. \ref{eq:mag}, one obtains \cite{Lem08}:
\begin{equation}
 L_{\mathrm {B}} \ge \frac{1}{4} (\eta Ze)^{-2} \theta^2 \Gamma^2 \beta^3 c E_{\mathrm {obs}}^2
\label{eq:lumb}
\end{equation}
As discussed previously, the jet magnetic power is not directly observable, however this form of
energy, probably dominant at the beginning, transforms into kinetic one via propagation-related 
processes. Most importantly, jet kinetic power has been connected to the (observable) luminosity
of the radio core by
\cite{All06} and \cite{Hein07} who derived this quantity from the work done
by the radio lobes to evacuate intracluster cavities, seen at the X-ray energy band. They showed
a correlation between the kinetic luminosity and the luminosity of the radio core, in
the limit of flat power-law spectra of the radio emission ($\alpha=0$):
\begin{equation}
 L_{\mathrm {kin}} = L_{\mathrm {k0}} \left(\frac{L^{\mathrm {core}}_\nu} {L_{\nu 0}} \right)^{12/17}
\label{eq:kinradio}
\end{equation}
With $L_{\nu 0}=7 \times 10^{22}$ W Hz$^{-1}$ and $L_{\mathrm {k0}} =10^{37}$ W.
This correlation is shown in the log-log plot of Fig. \ref{fig:heinz}, where the core radio power 
$L_{\mathrm {R}}=\nu  L^{\mathrm {core}}_\nu$, with $\nu=5$ GHz, and the luminosities are in cgs units.

We have now all the elements for relating condition (\ref {eq:lumb}) with quantities {\it that can
be actually observed}. In fact, recalling the above discussion about the transformation of
a Poynting-flux dominated jet into a kinetically dominated one, 
 from Eq. (\ref {eq:lumb}) and Eq. (\ref {eq:kinradio}) we derive:
\begin{equation}
 L_{\mathrm {k0}} \left(\frac{L^{\mathrm {core}}_\nu}{L_{\nu 0}}\right)^{12/17} \ge \frac{1}{4} (\eta Ze)^{-2} 
\theta^2 \Gamma^2 \beta^3 c E_{\mathrm {obs}}^2
\label{eq:obser}
\end{equation}
Solving for $L^{\mathrm {core}}_\nu$, and in case of cosmic rays of $E=10^{20}$ eV, we have:
\begin{equation}
 L^{\mathrm {core}}_\nu \ge 10^{24} \left(\eta^{-1} Z^{-1} \theta \ \Gamma \beta^{3/2} E_{20}\right)^{17/6} 
\, {\rm W} \  {\rm Hz}^{-1}
\label{eq:lnu}
\end{equation}
This limit can be checked against observation of $L^{\mathrm {core}}_\nu$. However, it can be more instructive
to gain a visual impression of this limit, for FR Is, by using Fig. \ref{fig:ledlow}. For doing this we must convert
the radio core luminosity into the total luminosity. As shown by \cite{Giov01}, this two quantities
are correlated, once corrected for Doppler effects on the radio flux. Following \cite{Giov01}, we
set a mean value of $\Gamma=5$ for this correction, obtaining:
\begin{equation}
\log L^{\mathrm {core}}_\nu = 0.62 \ \log L^{\mathrm {tot}}_\nu + 8.41 
\label{eq:lnutot}
\end{equation}
We may set for the jet velocity $\beta =1$, $Z=1$ for protons, and a typical opening angle of
radio jet is $\theta \approx 10^\circ$. One thus obtains:
\begin{equation}
 L^{\mathrm {tot}}_\nu \ge 7 \times 10^{24} \left(\eta^{-1} \theta_{10} \Gamma_{5} E_{20}\right)^{4.6} \,
{\rm W} \ {\rm Hz}^{-1}
\label{eq:lnu1}
\end{equation}
that should be compared with the data in Fig. \ref{fig:ledlow}, after having
decided which is the favored particle acceleration mechanisms in order to  set the
parameter $\eta$. 
Setting $\eta=1$, condition (\ref{eq:lnu1}) is represented in Fig. \ref{fig:ledlow} by the thick, horizonthal line.
We notice that $\eta$ cannot attain
values much smaller that unity, otherwise we run out of FR I sources suitable for particle acceleration.

We will be then able to look among the about 100 radio-loud AGNs within $100$ Mpc the ones that fulfill
condition (\ref{eq:lnu1}) for accelerating particles up to $E=10^{20}$ eV.

\section{SUMMARY}

We have reviewed the main properties of AGNs, their separation into radio-loud and radio-quiet 
sources and the implications of this classification on the problem of cosmic ray acceleration
from these objects. We have then examined, on quite general bases, the conditions for UHECRs production
and set observational limits to the radio luminosity of radio sources in order to be able to be
sources of UHECRs.


\begin{thebibliography}{27}
%
\bibitem{All06} S. W. Allen, R. J. H. Dunn, A. C. Fabian, G. B. Taylor and C. S. Reynolds, 
MNRAS 372 (2006), 21.
\bibitem{Bier87} P. L. Biermann and P. A. Strittmatter, Ap. J. 322 (1987) 643.
\bibitem{Bier97} P. L. Biermann, J. Phys. G: Nucl. Part. Phys. 23 (1997) 1.
\bibitem{Bland00} R. D. Blandford, PhST 85 (2000), 191.
\bibitem{Cesrky93} C. Cesarsky and V. Ptuskin, 23rd IRCR (University of Calgary) 2 (1993) 341.
\bibitem{FR74} B. L. Fanaroff and J. M. Riley, MNRAS 167 (1974) 31.
\bibitem{Gian06}  D. Giannios and H. C. Spruit, A\&A 450 (2006), 887.
\bibitem{Giov01} G. Giovannini, W. D. Cotton, L. Feretti, L. Lara and T. Venturi, Ap. J. 552 (2001), 508.
\bibitem{Giov07} G. Giovannini, M. Giroletti and G.B. Taylor, A\&A 474 (2007), 409.
\bibitem{Hein07} S. Heinz, A. Merloni and J. Schwab, Ap. J. 658 (2007), L9.
\bibitem{Hillas84} A. M. Hillas, Ann. Rev. A\&A 22 (1984) 425.
\bibitem{Kang96} H. Kang, D. Ryu and T. W. Jones, Ap. J. 456 (1996) 422.
\bibitem{Kov07} Y. Y. Kovalev, M. L. Lister, D. C. Homan and K. I. Kellermann, Ap. J. 668 (2007), 27.
\bibitem{Ledl96} M. J. Ledlow and F. N. Owen, AJ 112 (1996) 9.
\bibitem{Lem08} M. Lemoine, SF2A-2008: Proceedings of the French Society of Astronomy and 
Astrophysics Eds.: C. Charbonnel, F. Combes and R. Samadi (http://proc.sf2a.asso.fr, p.247).
\bibitem{Lyut07} M. Lyutikov and R. Ouyed, Aph 27 (2007), 473.
\bibitem{Mass03} S. Massaglia, Ap\&SS 287 (2003) 223.
\bibitem{Mass07} S. Massaglia, NuPhS 165 (2007), 130.
\bibitem{Mau07} T. Mauch and E. M. Sadler, MNRAS 375 (2007), 931.
\bibitem{Milg95} M. Milgrom and V. Usov, Ap. J. Lett. 449 (1995) L37.
\bibitem{NW2000}  M. Nagano and A.A. Watson, Rev. Mod. Phys. 72 No. 3 (2000) 689.
\bibitem{Prie07} M. A. Prieto, J. Reunanen, Th. Beckert, K. Tristram, 
N. Neumayer, J. A. Fernandez and J. Acosta, ASPC 373 (2007), 600.
\bibitem{Prot92} R. J. Protheroe and A. P. Szabo, Phys. Rev. Lett. 69 (1992) 2885.
\bibitem{Rach93} J. P. Rachen and P. L. Biermann, A\&A 272 (1993) 161.
\bibitem{Ros08} P. Rossi, A. Mignone, G. Bodo, S. Massaglia, A. Ferrari, A\&A 488 (2008), 795.
\bibitem{Sik05} M. Sikora, M. C. Begelman, G. M. Madejski and J. -P. Lasota, Ap. J. 625 (2005), 72.
\bibitem{Urry95} C. M. Urry and P. Padovani, PASP 107 (1995), 803.
\bibitem{Vietri95} M. Vietri, Ap. J. 453 (1995) 883.
\end{thebibliography}
\end{document}